\documentclass[journal=ancac3,manuscript=article]{achemso}

\usepackage{chemformula} 
\usepackage[T1]{fontenc} 
\usepackage[version=3]{mhchem} 
\usepackage[T1]{fontenc}       
\usepackage{graphicx} 
\usepackage{subcaption} 
\usepackage[separate-uncertainty = true, multi-part-units = single, list-units = single, range-units = single]{siunitx} 
\DeclareSIUnit\Molar{\textsc{m}}
\DeclareSIUnit\rpm{rpm}
\DeclareSIUnit\ppm{ppm}
\DeclareSIUnit\kbt{k_BT}

\usepackage[noabbrev,nameinlink]{cleveref}
\usepackage{multirow}
\usepackage{makecell}
\usepackage{colortbl}

\author{Yogesh Shelke}
\affiliation[Leiden University]
{Soft Matter Physics, Huygens-Kamerlingh Onnes Laboratory, Leiden University, PO Box 9504, 2300 RA Leiden, The Netherlands.}
\author{Fabrizio Camerin}
\affiliation[Utrecht University]
{Soft Condensed Matter, Debye Institute for Nanomaterials Science, Utrecht University, Princetonplein 1, 3584 CC Utrecht, The Netherlands.}
\author{Susana Mar\'in-Aguilar}
\affiliation[Utrecht University]
{Soft Condensed Matter, Debye Institute for Nanomaterials Science, Utrecht University, Princetonplein 1, 3584 CC Utrecht, The Netherlands.}
\author{Ruben W. Verweij }
\affiliation[Leiden University]
{Soft Matter Physics, Huygens-Kamerlingh Onnes Laboratory, Leiden University, PO Box 9504, 2300 RA Leiden, The Netherlands.}
\author{Marjolein Dijkstra}
\affiliation[Utrecht University]
{Soft Condensed Matter, Debye Institute for Nanomaterials Science, Utrecht University, Princetonplein 1, 3584 CC Utrecht, The Netherlands.}
\author{Daniela J. Kraft}
\email{kraft@physics.leidenuniv.nl}
\affiliation[Leiden University]
{Soft Matter Physics, Huygens-Kamerlingh Onnes Laboratory, Leiden University, PO Box 9504, 2300 RA Leiden, The Netherlands.}

\title{Flexible Colloidal Molecules with Directional Bonds and Controlled Flexibility}


\begin{document}

\begin{abstract}
Colloidal molecules are ideal model systems for mimicking real molecules and can serve as   versatile building blocks for the bottom-up self-assembly of flexible and smart materials. While most colloidal molecules are rigid objects, the development of colloidal joints has made it possible to also include conformational flexibility  into colloidal molecules. However,  their unrestricted range of motion does not capture the restricted motion range and bond directionality that is typical of real molecules.

In this work, we create flexible colloidal molecules with an \textit{in situ} controllable motion range and bond directionality by assembling spherical particles onto cubes functionalized with complementary surface-mobile DNA. We assemble colloidal molecules with different coordination number of spheres by varying the size ratio and find that they feature a constrained range of motion above a critical size ratio. Using  theory and simulations, we show that the particle shape together with the multivalent bonds create an effective free-energy landscape for the motion of the sphere on the surface of the cube. We  quantify the confinement of the spheres on the surface of the cube and the probability to change facet. We find that temperature can be used as an extra  control parameter to switch \textit{in situ} between full and constrained flexibility of these colloidal molecules.
These flexible colloidal molecules with temperature switching motion range can be  used to investigate the effect of directional, yet flexible bonds in determining their self-assembly and phase behavior, and may be employed as constructional units in microrobotics and novel smart materials. 
\end{abstract}

\bigskip

Colloidal molecules are excellent models for real molecules and can be used to study the influence of shape and bond directionality in self-assembly processes and phase behavior.\cite{van2003chemistry, poon2004colloids, glotzer_anisotropy_2007, li2011colloidal,duguet2011design, he2020colloidal, marin2022guiding} However, current colloidal molecules are often  rigid objects, while the functionality of many molecules such as polymers, intrinsically disordered proteins, and tRNA hinges on their ability to adapt their structure. This conformational flexibility enables lock-and-key interactions, as well as faster and more specific binding, and has been proposed to affect their diffusive motion.\cite{ mellado1988diffusion, betts_analysis_1999, dyson2005intrinsically, illien2017diffusion}

The development of colloidal joints has recently made the fabrication of flexible colloidal molecules through self-assembly possible.\cite{chakraborty2017colloidal, chakraborty2022self} These consist of solid  particles or liquid droplets that can bind particles such that they can still laterally move on their surface~\cite{van2013solid,chakraborty2017colloidal,mcmullen2018freely,rinaldin2019colloid}. 
Their intrinsic bond flexibility has been shown to enhance yields in self-assembly\cite{chakraborty2022self}. Furthermore, flexible colloidal molecules have been used to demonstrate that flexibility enhances diffusion\cite{verweij2020flexibility} and that flexible linear and ring-like structures follow Flory theory for polymers\cite{verweij2021conformations, verweij2022brownian, mcmullen2018freely}. Besides, they have great potential for fundamental studies of their phase behavior, to understand and fabricate reconfigurable materials,\cite{mcmullen2022self,mitra2022coarse} and to store information \cite{phillips2014digital}. 

However, current realizations of flexible colloidal molecules exclusively feature bonds with an unrestricted range of motion, whereas the bonds of real molecules are often constrained to a specific range of motion due to the orbitals underlying their intramolecular bonds. The restricted motion range thus provides molecules with bond directionality. 
These features have not been realized in colloidal molecules yet, but would be powerful not only in view of their ability to serve as model systems but also to create reconfigurable materials and functional devices with multiple stable configurations. So far, only a combination of gravitational confinement and steric hindrance of the bound particles could keep their positional order unchanged, but could not confine their motion.\cite{chakraborty2022self}

Here, we experimentally realize flexible colloidal molecules with controlled flexibility and directionality by exploiting solid colloidal joints with an anisotropic particle shape. We demonstrate the creation of flexible colloidal molecules with directional bonds and controlled flexibility by mixing colloidal cubes equipped with surface mobile DNA-linkers~\cite{van2013solid,rinaldin2019colloid,  chakraborty2017colloidal} with an excess number of spheres functionalized with complementary strands. The assembled colloidal molecules consist of cubes surrounded by spheres connected with multivalent bonds of DNA linkers. The anisotropic particle shape is the key ingredient because it guides the position and controls the number of attached particles. Furthermore, the cubic particle shape constrains the lateral motion of the attached particles due to an interplay between the spatially extended multivalent bond formed between the particles and the curvature of the cube, thereby realizing controlled flexibility.
We show both by experiments and simulations that the combination of the cubic shape with the extended bonding patch provides confinement of the outer particles above a critical size ratio of sphere-to-cube diameter. We justify the observed behavior by calculating the free energy of the spheres experienced at different locations on the surface of the cube. The sensitivity of the DNA-based bonds to temperature allows us to lift confinement of the spheres to the cube sides \textit{in situ} by a simple elevation of temperature. 
Small colloidal molecules with controlled flexibility and directionality can be separated using a magnet due to the permanent magnetic dipole moment of the hematite cube, providing an easy and efficient means for their exploitation. The novel features of bond directionality and temperature-controlled switch from limited to full flexibility make these colloidal molecules excellent model systems for studying the phase behavior of molecules and building blocks. This would also enable the assembly of reconfigurable structures with multiple stable configurations, a key ingredient for creating functional devices and machines.  

\section{Results and Discussion}

\subsection{Assembly of Colloidal Molecules with Controlled Flexibility and Directionality}

\begin{figure}[p!]
	\centering
	\includegraphics [width=0.94\textwidth]{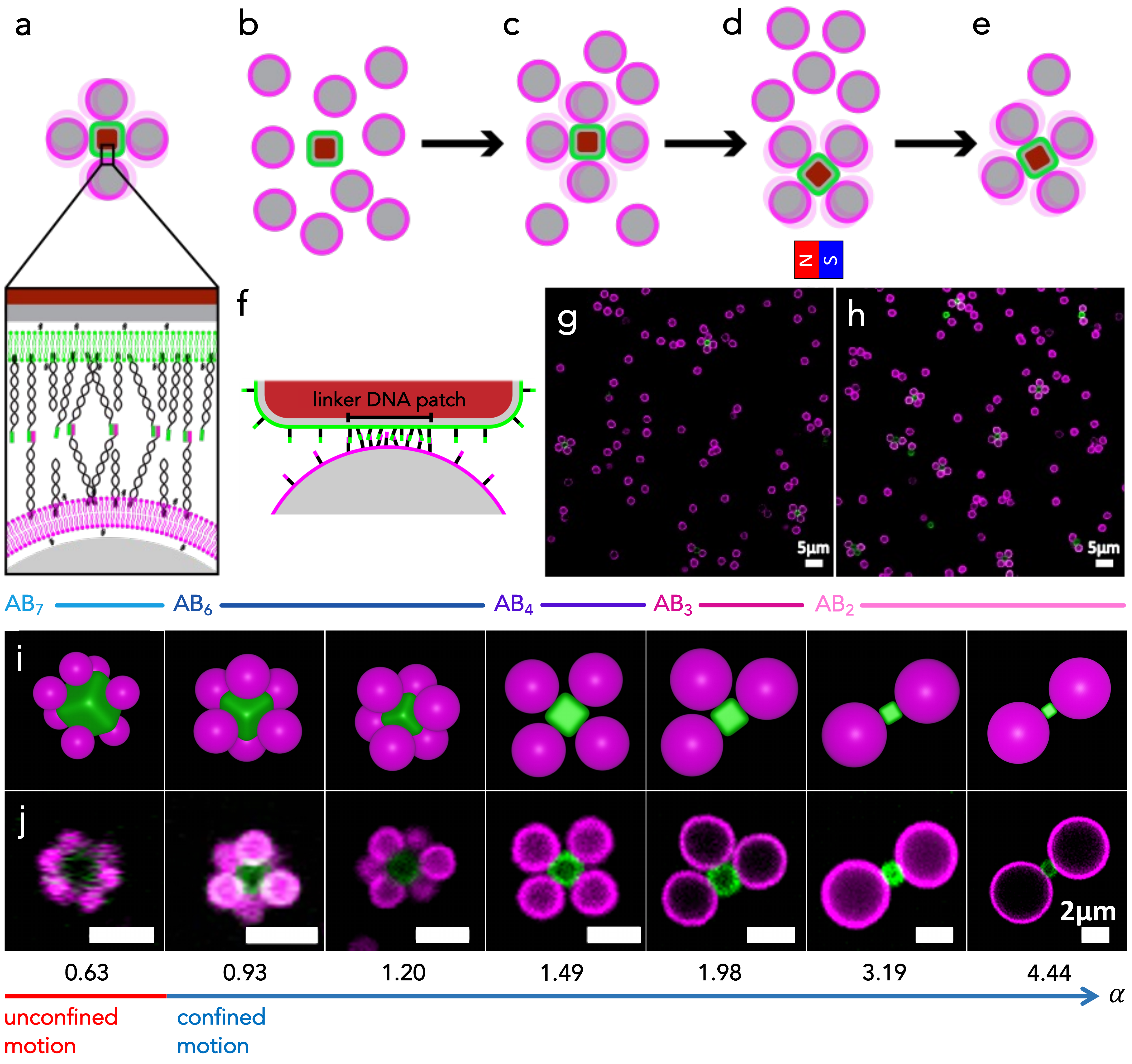}
	\caption{\small \textbf{Self-assembly of colloidal molecules with controlled flexibility and bond directionality}. (a) Schematic of a flexible colloidal molecule consisting of spheres (magenta) assembled onto a cube (green) and close-up view showing their functionalization with a lipid bilayer and surface-mobile DNA strands, where colored ends indicate different linker sequences. (b) The functionalized cubes and spheres are combined at a ratio of 1:20, which (c) enables rapid saturation of the cubic particles and ensures the formation of colloidal molecules. (d) The permanent magnetic dipole moment of the cubic particles is used to isolate the colloidal molecules by placing a magnet near the sample and removing the excess spheres, leading to (e) isolated flexible colloidal molecules. (f) Schematic indicating the patch formed between DNA linkers. (g) and (h) shows representative confocal microscopy images of assembled flexible colloidal molecules prior to and following isolation. (i) Schematics of the AB$_N$ assembled clusters and (j) representative confocal microscopy images of the most likely type of experimentally obtained colloidal molecules for varying sphere-to-cube size ratios $\alpha=0.63, 0.93, 1.20, 1.49, 1.98, 3.19$, and $4.44$ (from left to right). For $\alpha=0.63$, the spheres move indistinctly on different faces of the rounded cube, while an increasingly confined motion on single faces occurs for higher $\alpha$.
}
\label{fig:assembly}
\end{figure}

We create colloidal molecules with controlled flexibility and directionality by assembling spherical silica particles onto the surface of cubic colloids, see Figure \ref{fig:assembly}. 
We used solid silica spheres in combination with rounded cubic particles (also named superballs) made of hematite \cite{sugimoto1992preparation} and coated with a thin silica layer by a St\"ober procedure.\cite{wang2013shape} 

To achieve flexible bonds, we exploited a method previously developed by some of us  with modifications~\cite{van2013solid, chakraborty2017colloidal, rinaldin2019colloid}: the particles were functionalized with a lipid bilayer that contained dsDNA strands with and without a 11 bp single stranded end that act as linkers and steric stabilizers, respectively, see Figure \ref{fig:assembly}a. Since colloids coated with mobile linker DNA are able to bind more quickly to surfaces with higher receptor density\cite{linne2021direct,scheepers2020multivalent}, we here employed 300 strands/$\mu$m$^2$ on the spherical and 20000 strands/$\mu$m$^2$ on the cubic surfaces. After mixing, the complementary functionalized cubes and spheres bind to each other by accumulating DNA linkers in the area of closest contact. In this way, a so-called DNA \textit{patch area} is formed, as shown in Figure~\ref{fig:assembly}f.
The low concentration of DNA strands on the spheres limits the number of bound DNA linkers and in this way ensures mobility of the colloids after binding \cite{chakraborty2017colloidal}. The high linker concentration on the cubes provides fast binding of the spheres, even for the consecutive adsorption of spheres, because linker accumulation in the bond area does not lead to significant depletion of linkers on the cube surface.\cite{linne2021direct, verweij2021} In the Methods section, we report further experimental details and the DNA sequences employed.

In a typical experiment, we mixed the DNA-functionalized cubes with an excess number of spheres (cube-to-sphere number ratio 1:20) and transferred them to a customized sample holder, where polyacrylamide(PAA)-coated coverslips were used as a substrate and cover on the top. During self-assembly, the excess number of spheres and the high concentration of linker DNA on the cubes leads to faster binding of spheres onto the cubic particles and the formation of finite-size clusters, the flexible colloidal molecules, see Figure \ref{fig:assembly}b, c, and g.
After 12 hrs, the sample holder contained clusters and unbound spheres at the bottom. To remove excess spheres, we utilized the magnetic property of the hematite cube and isolated the flexible colloidal molecules by a handheld magnet. We did so by dispersing, separating and redispersing the colloidal molecules in buffer twice (200mM NaCl, 10mM HEPES, pH 7.4) (Figure~\ref{fig:assembly}d), which lead to a significantly increased purity and concentration of flexible colloidal molecules, see Figure~\ref{fig:assembly}h.

The cubic shape of the central particle allows for precise control over the maximum and most probable number of bound spheres by acting as a guiding template. The resulting colloidal molecules are of type AB$_N$, where A indicates the cube, B the spheres and $N$ the coordination number. We employed different size ratios of the sphere diameter $\sigma_s$ to cube edge length $\sigma_c$, $\alpha=\sigma_s/\sigma_c$= 0.63, 0.93, 1.20, 1.49, 1.98, 3.19, and 4.44. In this way, we find different values of the number of spheres attached to the cube, with more than six spheres attached for the lowest size ratios and with at maximum one sphere per facet for $\alpha \geq 1.20$. Schematic representations and confocal microscopy images of the resulting flexible colloidal molecules are shown in Figure~\ref{fig:assembly}i and j for the respective size ratios.  We note that magnetic separation works well for size ratios from 0.63 to 1.49, but for larger size ratios from 1.98 to 4.44, we found that the increased size of the sphere caused the bond between the sphere and cube to break.

The most probable coordination number of the colloidal molecules is the result of two factors. First, it is based on packing considerations, according to which each bounded sphere limits the available space for others to bind depending on its excluded volume.\cite{chakraborty2022self} This effect has been extensively explored in terms of yield and cluster size distribution for flexible colloidal molecules made of spheres only,\cite{chakraborty2022self} and for electrostatically assembled rigid colloidal molecules made from spheres and cubes~\cite{shelke2023exploiting}. For the latter, colloidal molecules with coordination number 6, AB$_6$, were found for sphere-to-cube size ratio $\alpha < 2$, while AB$_4$ and AB$_2$ type colloidal molecules were predominantly observed at around $\alpha \approx 3$ and for $\alpha > 3$, respectively. For the present case of flexible colloidal molecules, the cubic particle at the center provides a similar template that guides the position of the spheres. Indeed, for $\alpha \geq 1.20$ we find approximately the same coordination numbers, as evidenced by the most likely cluster shown in Figure \ref{fig:assembly}i and j. 
Differently from rigid colloidal molecules, we find that longer times are typically required for establishing the flexible bonds between DNA linkers, and hence we allowed assembly to continue for 12 hours instead of the 15 minutes that are typically required for rigid colloidal molecules.\cite{shelke2023exploiting}
We also found that employing a high number of DNA linkers on cubic particles increases the yield of colloidal molecules with maximum valence. The experimental cluster size distribution for the analyzed size ratios is reported in Figure S1.

The coordination number of the colloidal molecules is also influenced by the gravitational height which effectively confines larger spheres to quasi-2D, see Table S1. The size of the cubic particles cannot be varied largely, and hence different size ratios are achieved by varying the size of the spheres. 
For low $\alpha=0.63$, 0.93 and 1.20, we can employ small spheres which diffuse in three dimensions. Therefore, assembly occurs mainly in three-dimensional conditions and yields predominantly AB$_6$-type colloidal molecules. 
For higher size ratios which require larger spheres we mainly retrieve two-dimensional constructs, as the spheres quickly settle to the bottom of the container and assembly occurs under quasi-2D conditions. Hence, as at most only one sphere binds per facet, AB$_4$ and AB$_3$ colloidal molecules are obtained at $\alpha = 1.49$ and 1.98, respectively. Further increasing the size ratio above $\alpha = 3.19$ forces the spheres to bind to opposite sides of the cube due to steric constraints, forming mostly AB$_2$ colloidal molecules.

\subsection{Impact of Size Ratio on the Flexibility of the Colloidal Molecules}

The cubic shape in combination with the multivalent bonds is furthermore crucial for realizing the spatially constrained motion of the adhered spheres and the conformational flexibility of the colloidal molecules. As mentioned before, the bond between spheres and cubes is made up of many bound DNA linkers inside a patch area (see Figure~\ref{fig:assembly}f), whose size and shape are determined i) by the distance at which two DNA linkers can still bind and ii) by the shape, dimension and relative position of the two bound particles. Motion of the spheres on the surface of the cube implies a compaction at the edges and corners or an extension of the DNA patch on the facets, as previously observed for colloidal joints. \cite{chakraborty2017colloidal} Since the energy required for such rearrangement might be larger than the thermal energy or even require breaking bonds to sterically make patch compaction possible, we can expect crossing around a corner or edge of the cube to occur less frequently. Therefore, besides having access to different coordination numbers, by varying the size ratio between the spheres and the cube, and thus the features of the DNA patch area, we expect to be able to control the conformational flexibility of colloidal molecules. Eventually, we will be able to observe a transition from an unconstrained motion, where spheres are able to move across different facets, to a constrained motion on a single facet of the rounded cube.

\begin{figure}[t!]
	\centering
	\includegraphics [width=1\textwidth]{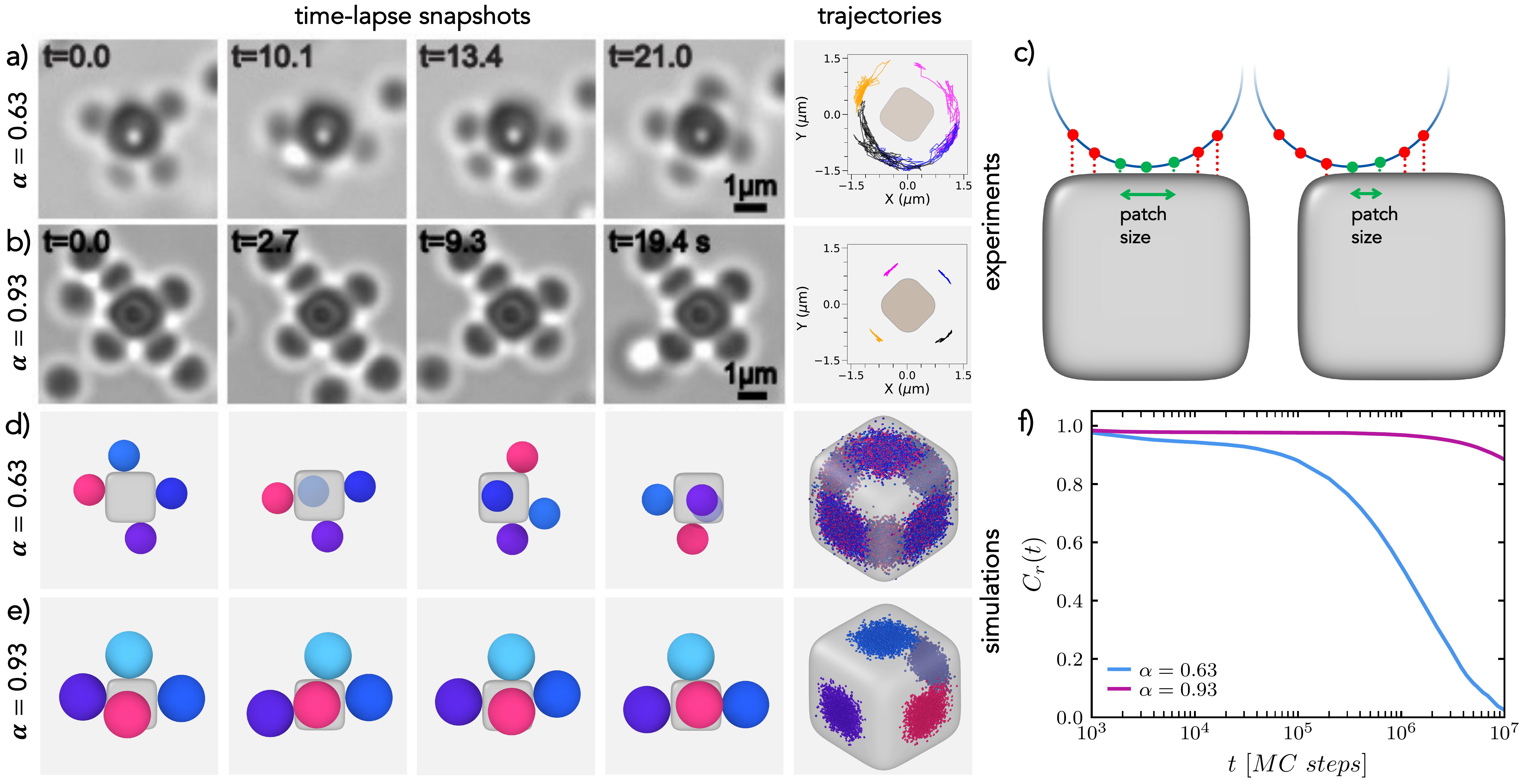}
	\caption{\small \textbf{Transition from unconstrained to constrained motion.} Time-lapse snapshots of microscope images showing the sphere motion on the cube for (a) $\alpha=0.63$ (unconstrained motion) and (b) $\alpha=0.93$ (constrained motion). The rightmost panel reports the trajectories of the spheres on the surface of the cube. (c) Schematics of the model used in simulation showing a portion of the tessellated sphere on the surface of the rounded cube. Depending on their relative position, the number of monomers interacting with an attractive interaction (green) and those not interacting (red) changes, implying a variation in the effective size of the patch area. (d,e) Three-dimensional time-lapse simulation snapshots and trajectories for the same size ratios as in (a) and (b), respectively. For each $\alpha$, the snapshots always have a fixed orientation and particles with the same identifier are colored alike. The rounded cube always has the same dimension in the two cases. (f) Position autocorrelation function $C_r(t)$ of the spheres as a function of Monte Carlo steps for the two size ratios $\alpha$ analyzed.}
	\label{fig:confinednotconfined}
\end{figure}

For analyzing these aspects, we first focus on two size ratios, namely $\alpha=0.63$ and $0.93$, and a colloidal molecule with four spheres attached. We intentionally introduced damage to the PAA coating to (partially) immobilize the flexible colloidal molecules by adhesion to the glass slide and subsequently selected one with an immobilized cube and mobile spherical particles (see Methods). In this way, it is possible to easily track the motion of the spheres on the central cubic particle and the results for the two different size ratios are not affected by excluded volume effects.
Figure~\ref{fig:confinednotconfined}a and b report the experimental time-lapse microscope images and the trajectories of the spheres over time for both size ratios $\alpha$. For the smallest size ratio $\alpha=0.63$, we observed that the spheres move smoothly between the different faces of the cube, implying that they were able to cross its edges. Correspondingly, the trajectories of the attached spheres show how the spheres are able to diffuse on the surface of the cube giving rise to an unconstrained motion. This picture drastically changes when studying $\alpha=0.93$. In this case, the motion is restricted to the facets of the cube, with the spheres always exploring the same facet where they initially bound to, as evident from the particle trajectory shown in the rightmost panel of Figure~\ref{fig:confinednotconfined}b. Further insights can be gained by studying the motion of the spheres onto the rounded cube by means of Monte Carlo simulations.

To this aim, we resort to a simplified model, according to which spheres are treated as tessellated objects and the central cubic particle as a superball. The model is schematically depicted in Figure~\ref{fig:confinednotconfined}c. Each point on the surface of a sphere interacts attractively with the cube when the reciprocal distance is comparable to the length of the DNA linkers in the experiments. In this way, we are able to account in a coarse-grained fashion for the rearrangements of the DNA patch when the relative position between the sphere and the cube changes. In fact, when the sphere is located in the center of the face the amount of monomers on the spheres favorably interacting, and thus the size of the DNA patch, is greater than when the sphere slides towards the edges or corners of the cube (see Figure~\ref{fig:confinednotconfined}c). Accordingly, moves towards such locations of the cube will occur with a lower probability as a result of the increase in energy that would be necessary. The strength of the attraction between monomers and cube is chosen according to the experimentally observed behavior and is fixed for all size ratios. For simplicity, we fix the monomer surface density of the spheres and the curvature of the rounded cube, which are likely to influence the motion of the spheres in various ways. Further details on the modeling and simulations are reported in the Methods and in the Supplementary Material. 

Using the same size ratios as in the experiments, we indeed recover a similar behavior. Simulation snapshots and the corresponding regions of the cube explored by the four spheres are shown in Figure~\ref{fig:confinednotconfined}d and e for $\alpha=0.63$ and $0.93$, respectively. Such regions are made up by the contact points of the spheres onto the superball and are colored differently for different spheres. In agreement with experiments, we observe that the increase in size ratio from $\alpha=0.63$ to $\alpha=0.93$ corresponds to a transition from unconstrained to constrained motion. Indeed, we observe that small spheres with $\alpha=0.63$ change multiple times between facets, implying that their movement is free on the surface of the cube and across faces (see Figure~\ref{fig:confinednotconfined}d).
 
This behavior is also reflected in the decay of the position autocorrelation function, defined as $C_r(t)=\sum_{i=1}^N (1/N) ({\bf r}_i(0) \cdot {\bf r}_i(t))/{\bf r}_i(0)^2$, where the sum runs over all bounded particles $N=4$, ${\bf r}_i(0)$ and ${\bf r}_i(t)$ are the initial position and the position at time $t$, respectively, of the $i$th sphere. Note that the positions are taken as the contact point of the sphere with the surface of the cube. For $\alpha=0.63$, $C_r(t)$ starts already decaying after $10^4$ MC steps, implying that particles indeed diffuse freely on the surface of a cube, see Figure~\ref{fig:confinednotconfined}f. On the contrary, for $\alpha=0.93$, we confirm the confined movement on the initial faces of adsorption, with $C_r(t)$ not decaying even at long times.

\begin{figure}[t!]
	\centering
	 \includegraphics [width=1\textwidth]{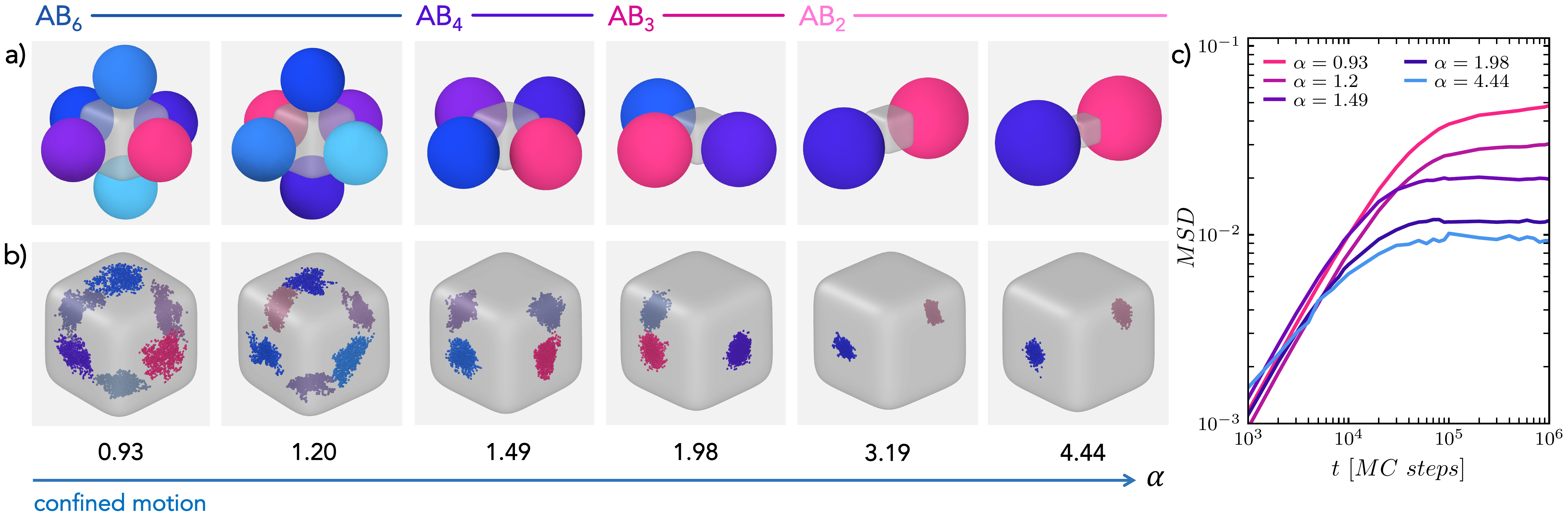} 
	 \caption{\small \textbf{Confined motion.} (a) Simulation snapshots showing an exemplary configuration of the spheres with respect to the central cube and (b) trajectories of the points of contact of the bound spheres for varying sphere-to-cube size ratios $\alpha=0.93, 1.20, 1.49, 1.98, 3.19$, and $4.44$ (from left to right). For better visualization, the points of contact are reported every fifty Monte Carlo (MC) steps. Different colors are employed to distinguish different spheres. (c) Mean-squared displacement $MSD$ of the spheres as a function of MC steps for the same size ratios $\alpha$ as in (a,b).
	 }
	\label{fig:constrainedunconstrained_sim}
\end{figure}

The study of the dynamics of the spheres on the surface of the rounded cube can also be extended to colloidal molecules with a sphere-to-cube size ratio that exceeds $0.93$.
In simulations, we assess the most frequently experimentally assembled cluster for each size ratio $\alpha$ and we constrain the motion of the spheres in the $z$-direction to mimic the effect of the experimentally observed quasi two-dimensional confinement. Configurations adopted by such colloidal molecules and particles trajectories are shown in Figure~\ref{fig:constrainedunconstrained_sim}a and b, respectively. The corresponding experimental frames are shown in Figure~\ref{fig:assembly}j.
Consistently with the previous observation for $\alpha=0.93$, we notice that the sphere motion is always restricted to the same facet.  Nonetheless, these colloidal molecules with limited mobility of the spheres still retain a discrete amount of flexibility, which in these cases thus refer to the ability of the spheres to explore the face they are bound to. While all colloidal molecules indeed possess some flexibility, the range of motion of the attached spheres reduces with increasing size ratio. This can be observed by looking at the regions explored by the spheres reported in Figure~\ref{fig:constrainedunconstrained_sim}b, where the scattering of the sphere's contact points is progressively reduced for higher $\alpha$. This phenomenon is due to the increased patch size of the bound DNA linkers with respect to the cube size, which implies that the spatial limitation by the cube facet is already experienced  at smaller displacements from the center of the facet. Thus, with increasing $\alpha$, a further decrease of the flexibility is observed. Besides, steric constraints imposed by the presence of spheres on adjacent facets of the cubes can further reduce the range of motion at higher size ratios for the same coordination number $N$. The colloidal molecules thus obtained not only differ in coordination number but also in conformational flexibility. 
We quantify the confinement in the motion by calculating in simulations the mean-square displacement of the spheres  $\sum_{i=1}^N  |{\bf r}_i(t)-{\bf r}_i(0)|^2/N$, where we assumed the surface area in which the sphere motion occurs to be flat. We report the latter as a function of Monte Carlo steps in Figure~\ref{fig:constrainedunconstrained_sim}c for the different size ratios studied. The flattening of the mean-square displacement at long times confirms the presence of constrained motion which is progressively restricted towards $\alpha=4.44$. In particular, we note that the average area explored by the spheres on the faces, taken as the plateau value of the mean-square displacement, is restricted by more than six times with increasing $\alpha$.

\subsection{Origin of Constrained and Unconstrained Motion of Flexible Colloidal Molecules}

\begin{figure}[t!]
	\centering
	 \includegraphics [width=1\textwidth]{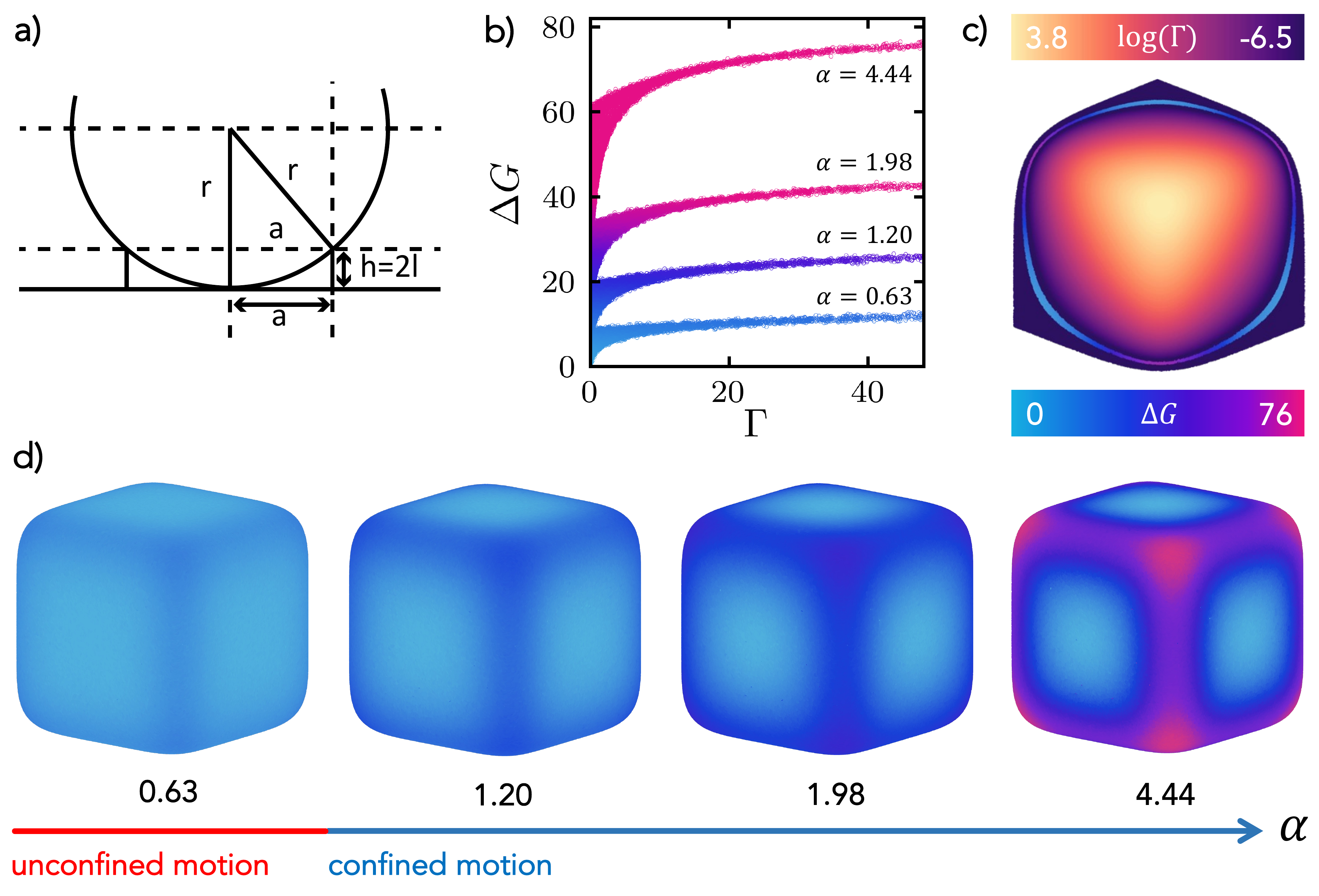} 
	\caption{\small \textbf{Origin of confined and unconfined motion.} (a) Schematic illustrating the patch size between sphere and cube. (b) Gibbs free energy $\Delta G$ as a function of the Gaussian curvature of the superball $\Gamma$ for the four different size ratios analyzed, namely $\alpha = 0.63, 1.20, 1.98$ and $4.44$. (c) Octant of the superball depicted according to the Gaussian curvature. For a band of the octant with low curvature, we also report the corresponding Gibbs free energy. For the other octants, the free energy landscape is the same due to the symmetry of the superball. (d) Gibbs free energy $\Delta G$ reported on the surface of the superballs for the four different size ratios analyzed in (b). The color scale is the same as in (c).}
	\label{fig:freeenergy}
\end{figure}

The model employed in the previous section allowed us to provide an effective description of the particle mobility on the surface of the rounded cube. We now proceed to identify the nature of the constrained and unconstrained motion, by first estimating the patch size on the edges and faces of the cube and then by calculating the different contributions to the free energy. For a uniform DNA coverage, the size of the DNA patch formed between a sphere and a cube depends on the position of the point of contact. It scales with the size of the sphere and can be calculated according to the scheme shown in Figure \ref{fig:freeenergy}a, where the extension of the patch area is determined by the maximum distance at which two linkers can form a bond. While a sphere can initially bind either at a rounded corner, edge or a flat face of the cube, it will preferentially end up being bound to the flat faces of the rounded cubes, because there the patch will possess its maximum extension. 
Assuming a flat face, this corresponds approximately to 
$A_{face}=2\pi h(2r-h)$,
where $h$ is the total length of both DNA linkers and $r=\sigma_s/2$ the radius of the spherical colloid. This picture changes when a sphere approaches the edge of a cube and the surface is curved. There, the patch will not be able to maintain its initial conformation but has to rearrange and compress before attempting to cross from one face of the cube to another one. We remark that this process does not necessarily imply any breaking of the DNA links, but their rearrangement due to the mobility with which they are endowed by being embedded in the lipid bilayer. Approximating the rounded edge by a sphere of radius $R$, the patch area that can be formed when the sphere is precisely located at the center of the edge is given by $A_{edge}= 2 \pi R \left( R-\sqrt{R^2-\frac{4l(R-l)(r-l)(R+r-l)}{(R+r-2l)^2}} \right)$, where $R$ denotes the cube edge curvature radius and $l$ is the length of DNA linker.\cite{chakraborty2017colloidal}
For $\alpha = 0.63$, the DNA patch size is  0.11 $\mu$m$^2$ on the cube's facet and 0.10 $\mu$m$^2$ on the cube edge, whereas for $\alpha = 0.93$, the DNA patch size reduces significantly from the flat face to the edge, from 0.18 $\mu$m$^2$ to 0.07 $\mu$m$^2$. This affects the free energy of the sphere bound to different parts of the cube. 

Similarly to other processes where DNA-mediated interactions are involved, the change in Gibbs free energy $\beta \Delta G$ for moving from the center of the facet to another position on the cube can be written as~\cite{parolini2015volume,leunissen2011numerical}\begin{equation}\label{eq:deltag}
\beta \Delta G = \beta\Delta G^0 -\frac{\Delta S_{conf}}{k_B} ,
\end{equation}
 
where $\beta\Delta G^0$ accounts for the Gibbs free energy contribution for the formation of linkers between the DNA on the sphere and that on the cube, $\beta=1/k_BT$, $T$ is the temperature, $k_B$ Boltzmann's constant,  and $\Delta S_{conf}$ is the configurational entropy. 
The first term can be expressed as
\begin{equation}\label{eq:deltag0}
\beta\Delta G^0 = -\frac{\epsilon (N_{pos}-N_{face})}{k_BT} 
\end{equation}
where $\epsilon >0$ is the energy scale in our simulations (see Methods), and $N_{face}$ and $N_{pos}$ are the number of links formed between a cube and a  sphere  at the center of one of the faces of the cube and at a given position on the cube, respectively.
$\Delta S_{conf}$, instead, quantifies the loss in configurational entropy due to the rearrangement and compression of the patch.
This contribution is proportional to the area of the patch and can be written as
\begin{equation}\label{eq:deltasconf}
\beta\Delta S_{conf} \propto  \log \left( \frac{A_{pos}}{A_{face}}\right),
\end{equation} with $A_{pos}$ and $A_{face}$ a measure of the area of the DNA patch at a given position on the cube and at the face, where the extension is at its maximum, respectively. Both $\Delta G^0$ and $\Delta S_{conf}$ thus inherently depend on the size of the sphere. To quantify $\Delta G$,  
we make use of our simulation model depicted in Figure~\ref{fig:confinednotconfined}c and we calculate for each point on the surface of a cube the number of linkers and the patch area at that position. These are determined by accounting for the region of the sphere whose distance to the surface of the cube does not exceed twice the length of the DNA linkers (see Methods). Figure~\ref{fig:freeenergy}b reports the Gibbs free energy $\Delta G$ as a function of the Gaussian curvature $\Gamma$ of the rounded cubes for four different size ratios $\alpha$. $\Gamma$ is shown for an octant of a superball with $\alpha=4.44$ in Figure~\ref{fig:freeenergy}c~\cite{ni2012phase,torres2017general}. The Gibbs free energy is also depicted directly on the surfaces of the superballs in Figure~\ref{fig:freeenergy}d.

Consistently with our expectations, we find that the Gibbs free energy barrier for moving across the edges progressively grows as the diameter of the sphere is increased moving from $\alpha=0.63$ to $\alpha=4.44$, reflecting the higher cost for the rearrangement of the DNA patch. Therefore, the motion of the attached particle will be hindered and the probability to cross from one face to another will decrease for spheres with a larger diameter. Incidentally, the constrained motion on a given facet that we observed for increasing size ratio $\alpha$ (see Figure~\ref{fig:constrainedunconstrained_sim}) is captured by the fact that regions with the most favorable free energy are progressively reduced in size with increasing $\alpha$, as clearly shown in Figure~\ref{fig:freeenergy}d. 
Besides, for each size ratio, the highest value of free energy is found at the highest curvature, corresponding to the value measured at the corners of the rounded cube. 
We further notice that in all cases the system spans a range of free energies that is increasingly wider as $\Gamma$ tends to zero. This fact can be understood by looking at Figure~\ref{fig:freeenergy}c where, together with $\Gamma$, we also depict the free energies for a narrow band of the superball where the curvature has approximately the same (low) values. In this domain, we observe that moving from one face of the cube to the other we cross a region, corresponding to the edge, in which the free energy takes on radically different values. This counter-intuitive effect is related to the difference in area that the DNA patch experiences in these two regions, causing sites with similar curvature to have different free energies. In contrast, supposing we had an infinitely large surface with the same curvature, the free energy would have been the same.

We also note that a second, small, contribution to the scattering of the free energy values at the same curvature is likely to be intrinsic to our modeling, and specifically related to the way in which the spheres are tessellated. 
 
 \subsection{Temperature-Induced Reversible Transition of Flexible Colloidal \\ 
 Molecules from Constrained to Unconstrained Motion} 
 
The Gibbs free-energy difference experienced by the spheres hinges on the DNA-based bonding patch. While the maximum configurational entropy difference between the facet and edge is determined by geometry alone, the enthalpic and entropic contributions for the formation of the DNA patch depends on the number of bound DNA linkers via $\Delta G^0$ in Eq.~\ref{eq:deltag}. Therefore, a change in the number of bound DNA linkers in the patch tunes the free-energy landscape, providing an additional mean of controlling the conformational flexibility and eventually the confinement of the motion of the spheres. We experimentally realize this by exploiting the temperature dependent binding probability of the DNA linkers. Increasing the temperature close to the melting temperature reduces the number of bound linkers and hence increases the probability of a sphere to cross to other facets. 

 \begin{figure}[p!]
	\centering
	 \includegraphics [width=0.7\textwidth]{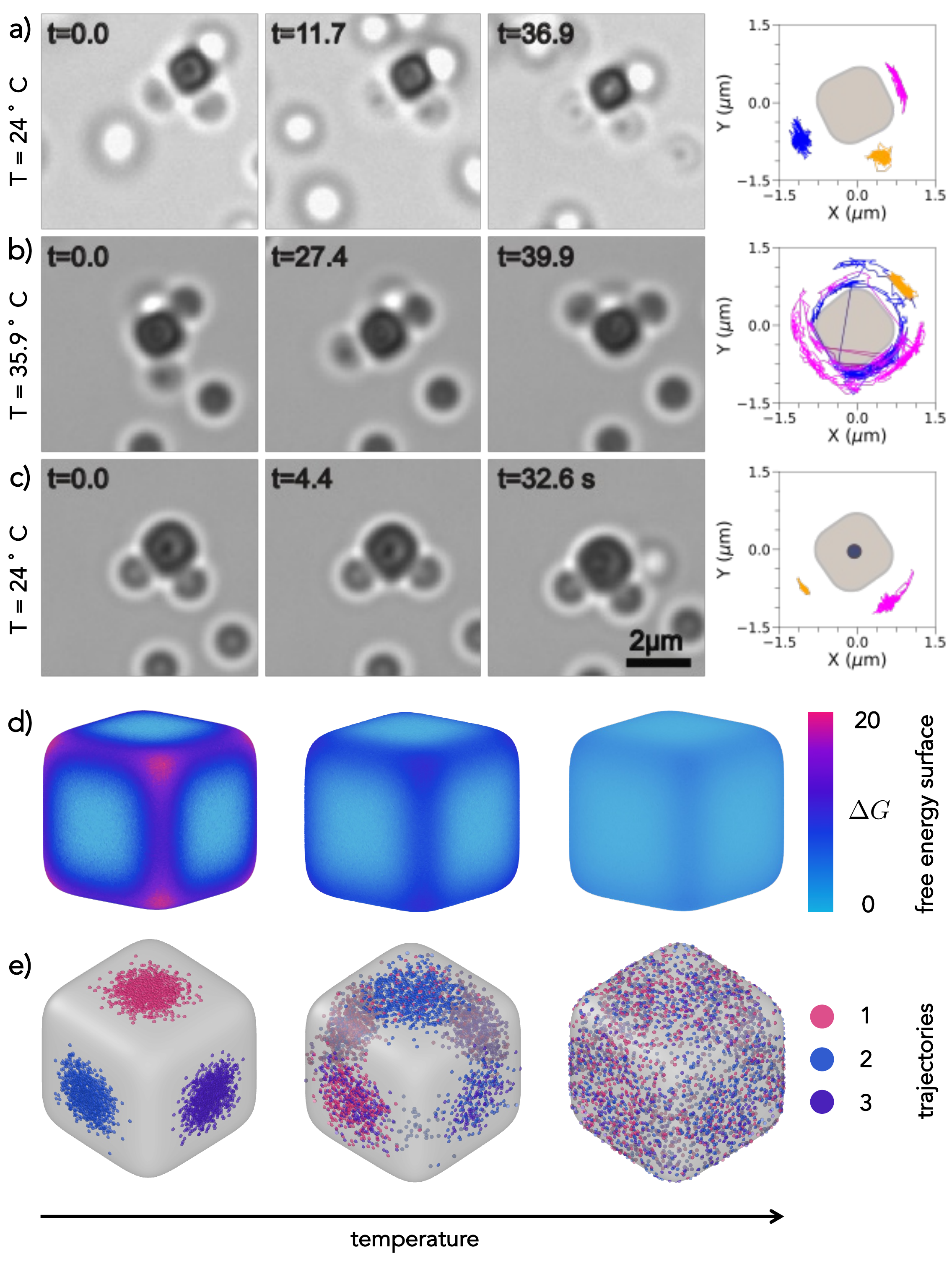} 
	 \caption{\small \textbf{Temperature-control over conformational flexibility.} (a-c) Bright field microscopy snapshots and corresponding trajectories of a flexible colloidal molecule with $\alpha=0.93$ showing a reversible transition of constrained to unconstrained conformational flexibility. The cube has been immobilized on the substrate. (a) At room temperature $T=24$ $^\circ$C the diffusion of spheres is constrained to the respective faces of the cube. (b) Upon an increase in the temperature to $T=35.9$ $^\circ$C the same flexible colloidal molecule shows full conformational flexibility where spheres can cross the edges. (c) Cooling back to room temperature confines the spheres' motion again to the respective faces of the cube. (d) Free energy $\Delta G$ reported on the surface of the superballs for three different effective temperatures (see Methods). (e) Trajectories of three spheres on the surface of the superball for increasing  temperatures for $\alpha=0.93$. Different colors relate to the trajectories of different spheres. For (d) and (e), the effective temperature increases from left to right.}
	\label{fig:temp-control}
\end{figure} 

We demonstrate the temperature dependent conformational flexibility with a flexible colloidal molecule for $\alpha=0.93$. Again, we immobilize the colloidal molecules to be able to track the particle motion (see Methods).
At room temperature, the spheres are mobile but confined to their respective faces as visible from the snapshots and trajectories of the spheres in Figure \ref{fig:temp-control}a. When the sample is heated to 35.9 $^\circ$C, the same flexible colloidal molecule now exhibits full conformational flexibility, and the spheres bound to cubes are able to move freely from one face of the cube to another. After the sample has been brought down to room temperature, the motion of the spheres once again is confined to the individual faces of the cube as seen from Figure~\ref{fig:temp-control}c. We note that one sphere remains confined to its side for the length of the video at  high temperature. This may be due to the sphere also being partially immobilized. Alternatively, it might stem from it having a higher DNA density, as the density of DNA linkers on a given sphere can vary by an order of magnitude\cite{chakraborty2017colloidal, linne2021direct} and hence even at higher temperature too many bonds may persist causing its motion to be constrained. Between the end of the cooling cycle and the start of the next one at lower temperature, the sphere achieved full mobility and became subsequently confined at the top of the cube. Therefore, only two traces are visible in Figure~\ref{fig:temp-control}c.  

We complement our experimental observations by also studying the effect of temperature on the conformational flexibility of the colloidal molecules with simulations (see Methods). 
For all cases, we report the Gibbs free energy and the particle trajectories on the surface of the cube in Figure~\ref{fig:temp-control}d and e, respectively. In Figure S3, the free energies are reported as a function of the curvature of the superball.
The free-energy calculations show that, by increasing the temperature, the free-energy differences towards the edge progressively decreases. Therefore, consistently with experiments, the crossing probability for spheres increases with temperature at the size ratio analyzed. This is the reason why at the highest effective temperature studied the three particles can freely explore the entire surface of the cube, as shown in the right panel of Figure~\ref{fig:temp-control}e. This phenomenon is similar to that observed earlier by decreasing the size ratio $\alpha$ between the sphere and the cube (see Figure~\ref{fig:confinednotconfined} and~\ref{fig:constrainedunconstrained_sim}). However, in this case, the change in free-energy barrier is entirely determined by the hybridization free energy $\Delta G^0$, while the configurational entropy term remains constant.

To summarize, the multivalent bonding patch together with the anisotropic shape of the cube provide therefore the necessary ingredients to confine the motion of the attached spheres and, by changing temperature, this constraint can be reversibly relieved and imposed on demand. 

\section{Conclusions}

We assembled colloidal molecules with directional bonds and controlled conformational flexibility by employing cubic particles at their center. By varying the size ratio $\alpha$ between the spheres and cubes, we assembled flexible colloidal molecules with different, well-controlled coordination number in high yields. We identified that the patch size of the bound DNA linkers between the sphere and the cube is critical in restricting the diffusive motion of the sphere to the cube's face. At $\alpha=0.63$, we find that spheres can easily diffuse across different faces of the cube, whereas their motion is constrained to a single face for $\alpha\geq 0.93$. The curvature variation of the cube leads to an effective free-energy landscape for the spheres' motion, with decreasing probability for crossing edges and corners with increasing size ratio. In addition, the motion on a given facet is increasingly confined due to the energetic costs associated with moving the patch to the more highly curved edges. We quantified their confinement and ability to change face, and analyzed their equilibrium distributions according to the free energies. We demonstrated that temperature can be used to reversibly switch between confined and unconfined motion of the spheres on the cubes' surface. The magnetic dipole moment of the hematite core of the cubic particles can be utilized to separate and clean the formed flexible colloidal molecules.

The thus prepared, flexible colloidal molecules can serve as building blocks for the preparation of higher order structures with desired flexibility and can be used to study the influence of controlled conformational flexibility on their phase behavior. Our insights into how the geometry of a template shapes the free-energy landscape for the adhered spheres are relevant beyond an application to flexible colloidal molecules as they can be employed to design other flexible structures with controlled conformational flexibility. The ability to release and reimpose the confinement by a simple increase and decrease of temperature is a powerful strategy to rearrange and fixate the conformation of flexible colloidal structures at will, thereby opening a way to create functional devices and machines.

\section{Materials and Methods}

\small
\subsection{Experimental Section}
\subsubsection{Materials}
Silica microspheres of diameters 0.97$\pm$0.05 $\mu$m,  1.25$\pm$0.05 $\mu$m, 1.55$\pm$0.05 $\mu$m, 2.06$\pm$0.05 $\mu$m, 3.32$\pm$0.05$\mu$m and 4.62$\pm$0.05 $\mu$m in 5 wt/v\% suspension were purchased from Microparticles GmbH. Silica particles of diameter 0.66$\pm$0.01 $\mu$m were synthesized by a St\"ober method. Sodium chloride, Ethanol, Sodium hydroxide, Ammoniun hydroxide (28-30 v/v\%),  Iron(III) Chloride Hexahydrate (FeCl$_3 \cdot $6H$_2$O), Tetraethyl orthosilicate (TEOS), 4-(2-hydroxyethyl)-1-piperazineethanesulfonic acid (HEPES), trimethoxysilyl propyl methacrylate (TPM), were purchased from Sigma-Aldrich. 1,2-dioleoyl-\textit{sn}-glycero-3-phosphocholine (DOPC), 1,2-dioleoyl-\textit{sn}-glycero-3-phosphoethanolamine-N-\-[methoxy(polyethylene glycol)-2000]  (DOPE-PEG2000), 1,2-dioleoyl-\textit{sn}-glycero-3-phospho\-ethanol\-amine-N-(lissamine rhodamine B sulfonyl)  (DOPE-Rhodamine) and dye 23-(dipyrrometheneboron difluoride)-24-norcholesterol (TopFluor-Cholesterol) were obtained from Avanti Polar Lipids, Inc.. We used Milli-Q water for all experiments. DNA strands were purchased from Eurogentec. The sequences of DNA used were: \\
Strand A: Double Stearyl-HEG-5$^\prime$-TT-TAT-CGC-TAC-CCT--TCG-CAC-AGT-CAC-CTT-CGC-ACA-GTC-ACA-TTC-AGA-GAG-CCC-TGT-CTA-GAG-AGC-CCT--GCC-TTA-CGA-\textit{GTA-GAA-GTA-GG-3$^\prime$-6FAM}, \\
Strand B: Double Stearyl-HEG-5$^\prime$-TT-TAT-CGC-TAC-CCT--TCG-CAC-AGT-CAC-CTT-CGC-ACA-GTC-ACA--TTC-AGA-GAG-CCC-TGT-CTA-GAG-AGC-CCT--GCC-TTA-CGA-\textit{CCT-ACT-TCT-AC-3$^\prime$-Cy3}, \\
Strand C: 5$^\prime$-TCG-TAA-GGC-AGG-GCT-CTC-TAG-ACA-GGG--CTC-TCT-GAA-TGT-GAC-TGT-GCG-AAG-GTG--ACT-GTG-CGA-AGG-GTA-GCG-ATT-TT-3$^\prime$, \\
Strand D: Double Stearyl-5TT-TAT-CGC-TAC-CCT-TCG-CAC-AGT-CAA-TCT-AGA-GAG-CCC-TGC-CTT-ACG-A  and Strand E: TCG-TAA-GGC-AGG-GCT-CTC-TAG-ATT-GAC-TGT-GCG-AAG-GGT-AGC-GAT-TTT. 

\subsubsection{Synthesis of Silica Cubes with Hematite Core.}
The hematite cubes of edge length 0.83 $\mu$m were  synthesized following reference\cite{sugimoto1992preparation} and coated with 0.105 $\mu$m silica layer using the process described in \cite{wang2013shape}. In a typical synthesis of hematite cubes, 100 ml of aqueous 2M FeCl$_3 \cdot $6H$_2$O was prepared in a 500 ml Pyrex bottle. Next, 100 ml of 5M NaOH solution were added while stirring for 20 seconds. Then, the mixture was stirred continuously for another 10 minutes and subsequently placed in a preheated oven and left undisturbed at 100 $^\circ$C for 8 days. The resulting hematite cubes were washed several times using centrifugation and redispersed in milliQ water. To coat them with a thin layer of silica, 100 ml of ethanol and 0.6 g of synthesized cubes were mixed under sonication and mechanical stirring in a 2-neck round bottom flask at 50 $^\circ$C. Subsequently, 5 ml of water, 15 ml of ammonium hydroxide solution and Tetraethyl orthosilicate (TEOS) were poured into the reaction flask. Next, the silica layer was allowed to grow on the cubes surface for 5h. The resulting particles were first washed with ethanol and then with water to remove unreacted chemicals by repeated centrifugation and redispersion. 

\subsubsection{Preparation of Single Unilamellar Vesicles (SUVs)}
Single unilamellar vesicles (SUVs) were prepared using a  protocol described in \cite{rinaldin2019colloid}. For the preparation of SUVs, we used 77 $\mu$l of 25 g/L DOPC, 7.34 $\mu$l of 10 g/L DOPE PEG 2000, and 2 $\mu$l of either 1 g/L dye DOPE-rhodamine or 2 $\mu$l of 1 g/L TopFluor-Cholesterol dissolved in chloroform were mixed together in glass vial. Subsequently, the lipid mixture was dried for at least 2 hrs in the desiccator (Kartell) attached with vacuum pump (KNF LABOPORT N816.3KT.18). Then, 1ml of buffer solution consisting of  50 mM NaCl and 10 mM HEPES at pH 7.4 was added to the dried lipid. The prepared solution was vortexed for 30 min during which the solution became turbid indicating  the formation of giant multilamellar vesicles. The dispersion of giant multilamellar vesicles was extruded (Avanti Polar Lipids mini extruder) 21 times through a 50 nm polycarbonate membrane supported with filter paper to achieve SUV formation. The prepared SUVs were stored in the fridge at 4 $^\circ$C and used for up to 3 days.  

\subsubsection{DNA Hybridization}
We used DNA strands with a hybridized backbone and a single stranded end (linker) and inert double-stranded DNA strands for bonding and stabilizing the colloids, respectively. Prior to use, single strand DNA was hybridized with the complementary backbone. Strand A was hybridized with strand C to yield double-stranded linker DNA , strand B with strand C to obtain complementary double-stranded linker DNA,  and strand D with strand E to create double-stranded inert DNA (DNA strands are listed in the materials section). For hybridization, we typically mixed 10 $\mu$l of 20 $\mu$M single strands and 10 $\mu$l of 20 $\mu$M complementary backbone in 90 $\mu$l buffer (200mM NaCl, 10mM HEPES, at pH 7.4) solution. The DNA solutions were placed in a preheated oven at 94 $^\circ$C for 30 minutes. The oven then was switched off and allowed to cool slowly overnight. After cooling, the hybridized DNA strands were stored at 4 $^\circ$C and used for up to 2 months.

\subsubsection{Functionalization of Colloidal Particles with a Lipid Bilayer Containing Linker and Inert DNA}
To coat particles with a lipid bilayer we used a 25:1 surface ratio of SUVs to particles. We maintained the same surface area ratio of SUVs to particles when coating differently sized particles. Typically, for 1$\mu$m particles, we use 100 $\mu$l of 0.25 wt/v\% particles in Milli-Q water and  mixed them with  48 $\mu$m 0.5 g/L  SUVs. Then, the dispersion was rotated at 8 rpm for 1h. During this period, SUVs collide, burst, spread and form a bilayer on the particles' surface.  Then, the coated particles were centrifuged at 800 rpm for 2-5 min and the supernatant containing excess SUVs was removed using a micropipette. Subsequently, a concentration of 300 linker DNA strands/$\mu$m$^2$ surface area was added to the spheres and 2×10$^4$ strands/$\mu$m$^2$ to the cubes. 1×10$^5$ strands/$\mu$m$^2$ of inert DNA was added to both particle suspensions. The suspensions were rotated for another 1h. Thereafter, each suspension was centrifuged and washed 2 times with a 50 mM NaCl buffer and then once with 200 mM NaCl buffer. The final suspension it was used in self-assembly experiments.  

\subsubsection{Sample Preparation, Magnetic Separation and Imaging} 
For all self-assembly experiments, a sphere-to-cube number ratio of 20:1 was maintained. In a 1.5 ml vial, 50 $\mu$l of functionalized 1 $\mu$m spherical particles and 2.5 $\mu$l of cubic particles were combined with 500 $\mu$l of buffer 2 solution. The mixture was then transferred to a customized sample holder and allowed to self-assemble for 12 h. Polyacrylamide (PAA) coated coverslips were used as a  substrate. The coverslips were coated with PAA by adapting the protocol in ref \cite{verweij2020flexibility}. 

For magnetic separation of the colloidal molecules after assembly, the sample holder was turned upside down to distribute the particles and colloidal molecules in 3D. Subsequently, a magnet was placed at the bottom of the sample holder to attract the flexible colloidal molecules while the non-magnetic excess spheres remained suspended in the sample. After 5 minutes, the supernatant containing the excess spheres was removed using a micropipette and the sample was resuspended in 200mM NaCl buffer solution. The same procedure was repeated one more time to further remove unbound spheres. 

To study how the conformational flexibility of the colloidal molecules changes with temperature, self-assembled $\alpha=0.93$ flexible colloidal molecules were deposited on coverslip with a damadged PAA coating. The PAA coating on the coverslip partially removed by scratching it with a fine needle. The sample was then placed on a custem-made microscpe stage attached with heating and cooling water circulater (JULABO DYNEO DD-300F) and temperature was monitored in the sample. The sample was heated in 1 $^\circ$C steps  and allowed to equilibrate for 10 min for each step during heating cycle and cooled linearly by circulating water through microscope stage.

Images and videos were captured with a  Nikon inverted TI-E microscope equipped with an A1  confocal scan head and a brightfield mode equipped with Prime BSI Express camera (Teledyne Photometrics). The images were taken using  with 100x oil objective (N.A. 1.4) at frame rates of up to 25 fps.

\subsection{Numerical Section}

Using simulations, we provide a coarse-grained description of the motion of the spheres, which are bounded on the surface of a rounded cube. The latter is part of the family of the so-called superballs which interpolate shapes between spheres and cubes and are described by:
\begin{equation}\label{eq:superball}
|x|^n+ |y|^n+|z|^n=\left|\frac{\sigma_c}{2}\right|^n,
\end{equation} 
where $n$ controls the roundness of the superball and $\sigma_c$ is the side of the superball that is also taken as the unit of length in our simulations. Here, we fix $n=6$, to capture the shape of the hematite cubes, while for $n=[2,\infty]$ we would retrieve all intermediate shapes between a sphere and a cube with sharp edges and corners. Given Eq.~\ref{eq:superball}, it is possible to define the Gaussian curvature at different positions on the surface of a superball as described for instance in Refs.~\cite{ni2012phase,torres2017general}.
The bounded spheres of diameter $\sigma_s$ are treated as tessellated rigid objects, with their surface covered by point-like monomers, as schematically shown in Figure~\ref{fig:confinednotconfined}c and Figure S2. By keeping fixed the size of the superball and varying the size ratio between the sphere and the rounded cube $\alpha=\sigma_s/\sigma_c$ between $0.63$ and $4.44$, the monomers surface density is kept approximately constant. In all cases, we always require the module of the normal vector connecting the surface of the cube to the center of the sphere to be $|\vec{n}| = \alpha\sigma_c$, thus enforcing the sphere to be located on the surface of the superball.

Monomers belonging to a bound sphere interact with the cube via a square-well potential 
\begin{equation}
\beta U_{cs}(s_{ij})=\begin{cases}
-\beta\epsilon & s_{ij}\leq \lambda\sigma_c \\
0 & s_{ij}>\lambda\sigma_c
\end{cases}
\end{equation}
where $\beta=1/k_BT$ with $k_B$ the Boltzmann constant and $T$ the temperature, $s_{ij}=|\vec{s}|$ denotes the distance between each point and the surface of the cube with $\vec{s}=s_{ij}\hat{n}$, $\epsilon=10 k_BT$ is the strength of the interaction potential and $\lambda\sigma_c=0.05\sigma_c$ its range, which is set to match the experimental length of DNA linkers. For mimicking the effect of a change in temperature, which experimentally affects the number of DNA linkers between spheres and superball, we also employ $\epsilon=1 k_BT$ and $\epsilon=5 k_BT$, corresponding to higher temperatures than the ones used for the main results in the first sections.

Finally, two bound spheres experience an infinite repulsive interaction at contact, while interactions between point-like monomers on the same sphere are not considered.

To study the motion of the bound spheres, we initialize each configuration by randomly placing, on the surface of a rounded cube, a number of spheres that is equivalent to the experimental average cluster size at the specific $\alpha$ analyzed.
 
We then perform Monte Carlo (MC) simulations by first letting the system to equilibrate for $10^5$ MC steps and then by recording trajectories for at least $10^7$ MC steps. To account for inhomogeneities in the tessellation of the bound particles, we rotate each sphere around a random axis for $10^2$ times every movement. In addition, we mimic the effects of the experimental gravitational height according to which, depending on the size of the bounded spheres, the assembly and the subsequent motion is restricted in space. Therefore, for size ratios $\alpha \geq 1.49$, the motion can only occur in quasi two-dimensional conditions, thus avoiding the upper and lower faces of the cube. No restriction is instead imposed for smaller $\alpha$. In all cases, the length of the displacement of each MC movement is calculated in such a way that the acceptance ratio is around $0.3$.

For the determination of the free-energy surface, we use the same model described above, we finely span the surface of the rounded cube, and for each point, we calculate the projected area of a bound sphere that is effectively linked to the cube. We do this by accounting for the monomers on the tesselated sphere that experience an effective attraction via the square-well potential. To account for inhomogeneities in the tessellation, we average the free energy for each point that we explore over  $10^3$ rotations of the sphere around a random axis.

\normalsize

\begin{acknowledgement} 
DJK gratefully acknowledges funding from the European Research Council (ERC Starting Grant number 758383, RECONFMAT). FC, SMA and MD acknowledge financial support from the European Research Council (ERC Advanced Grant number ERC-2019-ADV-H2020 884902, SoftML).
 
\end{acknowledgement}
\clearpage

\bibliography{ms}

\end{document}


\renewcommand{\thefigure}{S\arabic{figure}}
\renewcommand{\thetable}{S\arabic{table}}

\begin{table}[b!]
	\centering
	\caption{Gravitational height of the silica particles used in the study.}
	\label{height}
	\begin{tabular}{ccccc}
		\hline
	    \makecell{Diameter of Sphere\\ ($\mu$m)} & \makecell{Gravitational Height\\ (nm)}\\
		\hline
		0.66$\pm$0.01   & 2.8×10$^{3}$     \\
		0.97$\pm$0.05   & 8.8×10$^{2}$   \\
	    1.25$\pm$0.05   & 4.1×10$^{2}$  \\
	    1.55$\pm$0.05   & 2.2×10$^{2}$  \\
	    2.06$\pm$0.05   & 92 \\
	    3.32$\pm$0.05   & 22 \\
	    4.62$\pm$0.05   & 8.1 \\
	    \hline
	\end{tabular}
\end{table}

\begin{figure}[t!]
	\centering
	\includegraphics [width=1\textwidth]{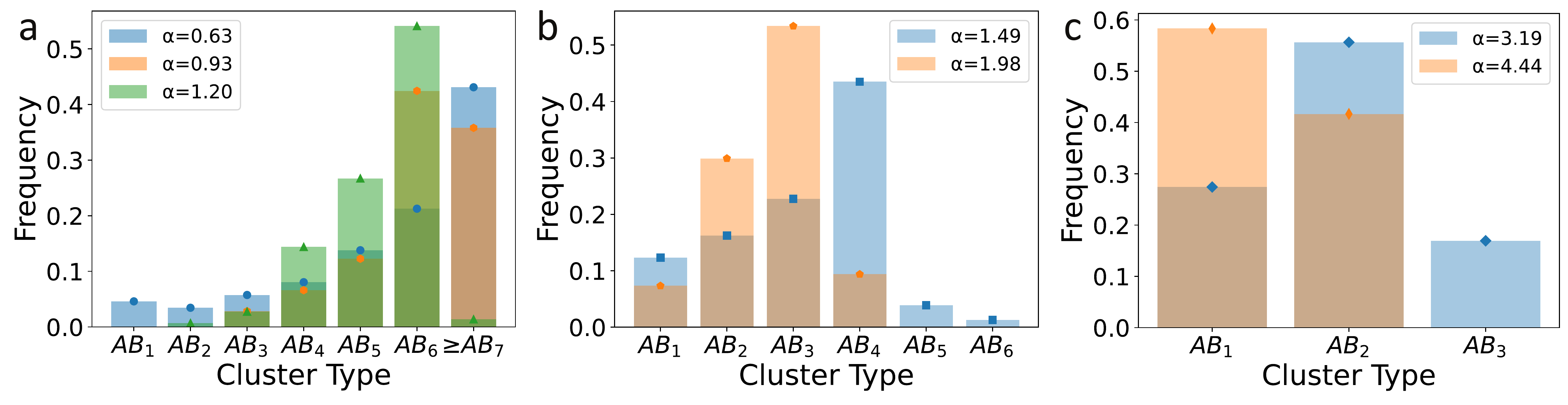}
	\caption{Cluster size distribution for varying sphere-to-cube size ratios a) $\alpha$=0.63, 0.93 and 1.20. b) $\alpha$=1.49 and 1.98, and c) $\alpha$=3.19 and 4.44. }
	\label{fig:cluster_size}
\end{figure}

\begin{figure}[b!]
	\centering
	\includegraphics[width=0.7\textwidth]{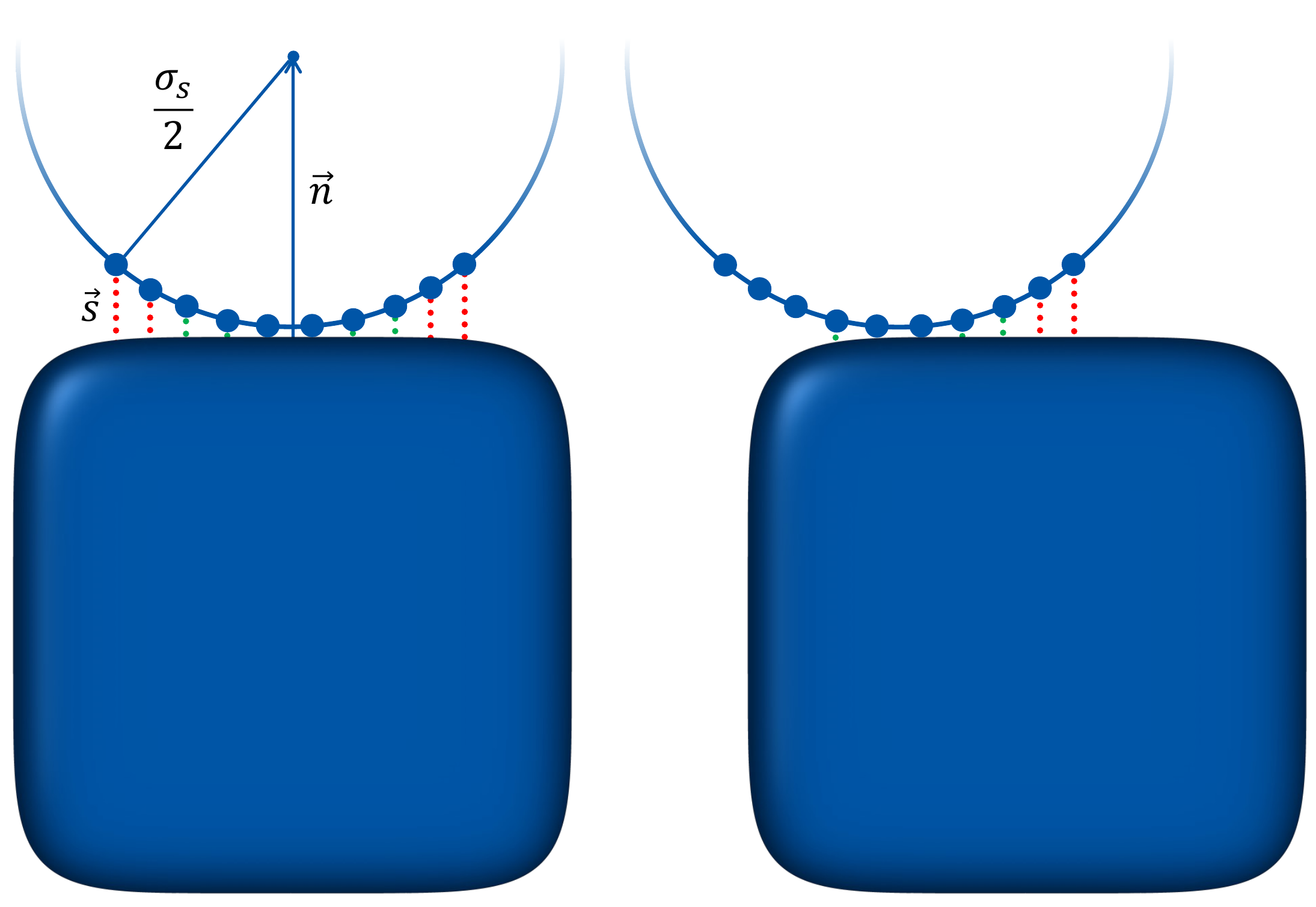}
	\caption{Schematics illustrating the numerical model, which consists of tessellated spheres of size $\sigma_s$ moving on the surface of a rounded cube of size $\sigma_c$. Two different relative sphere-cube positions are shown. Each point-like monomer on the sphere interacts with the cube only when the distance from its surface does not exceed the extension of the square-well potential $\lambda$ (see text). In here, green and red lines are for interacting or non-interacting monomers, respectively. The normal vectors $\vec{n}$ and $\vec{s}$ connect respectively the center of the sphere and point-like monomers to the surface of the rounded cube.}
	\label{fig:model}
\end{figure}

\begin{figure}[t!]
	\centering
	\includegraphics [width=0.8\textwidth]{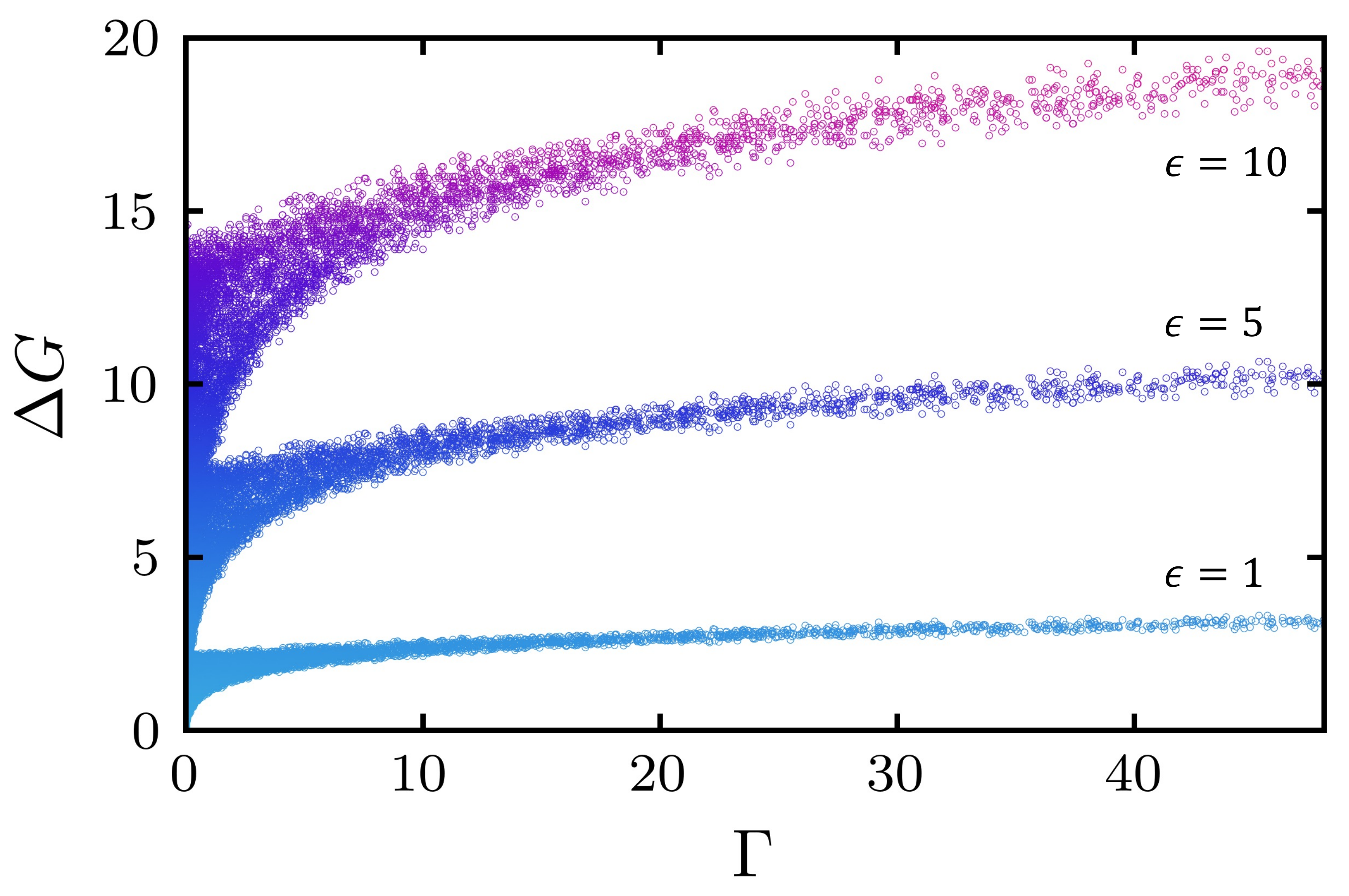}
	\caption{Free energy $\Delta G$ as a function of the Gaussian curvature of the superball $\Gamma$ for the three different effective temperatures analyzed, corresponding to $\epsilon=1, 5, 10$ (see Methods). }
	\label{fig:free energy}
\end{figure}